\documentclass[prb,aps,twocolumn]{revtex4}
\pdfoutput=1
\usepackage{epsfig}
\usepackage{epsf}
\usepackage{amsmath}
\usepackage{bm}

\newcommand{\be}{\begin{equation}}
\newcommand{\ee}{\end{equation}}
\newcommand{\bea}{\begin{eqnarray}}
\newcommand{\eea}{\end{eqnarray}}
\newcommand{\bei}{\begin{itemize}}
\newcommand{\eei}{\end{itemize}}

\begin{document}

\title{\bf Finite temperature phase transition for disordered weakly interacting bosons in one dimension}

\author{I.L. Aleiner$^{1}$, B.L. Altshuler$^{1}$, and G.V. Shlyapnikov$^{2,3}$}
\affiliation{\mbox{$^{1}$Physics Department, Columbia University, 538 West 120th Street, New York, New York 10027, USA}\\
\mbox{$^{2}$Laboratoire de Physique Th{\'e}orique et Mod{\`e}les Statistique,  Universit{\' e} Paris Sud, CNRS, 91405 Orsay, France}\\
\mbox{$^{3}$Van der Waals-Zeeman Institute, University of Amsterdam, Valckenierstraat 65/67, 1018 XE Amsterdam, The Netherlands}}

\begin{abstract}
\end{abstract}
\maketitle

  {\bf It is commonly accepted that there are no phase transitions in
    one-dimensional (1D) systems at a finite temperature, because
    long-range correlations are destroyed by thermal fluctuations.
    Here we demonstrate that the 1D gas of short-range interacting bosons
    in the presence of disorder can undergo a finite temperature phase
    transition between two distinct states: fluid and insulator.  None
    of these states has long-range spatial correlations, but this is a
    true albeit non-conventional phase transition because transport
    properties are singular at the transition point. In the fluid
    phase the mass transport is possible, whereas in the insulator
    phase it is completely blocked even at finite temperatures. We
    thus reveal how the interaction between disordered bosons
    influences their Anderson localization. This key question, first
    raised for electrons in solids, is now crucial for the studies of
    atomic bosons where recent experiments have demonstrated Anderson
    localization in expanding very dilute quasi-1D clouds.}



The absence of finite temperature phase transitions in one-dimensional
(1D) systems is considered as almost a dogma. Its justification is
based on another dogma which states that any phase transition is
related to the appearance/disappearance of a long-range order or at
least long-range spatial correlations.  Thermal fluctuations in 1D
systems destroy any long-range order, lead to exponential decay of all
spatial correlation functions and thus make phase transitions
impossible \cite{textbook1,textbook2}.  Non-interacting quantum particles in a
one-dimensional random potential show a similar behavior in the sense
that all single-particle eigenfunctions are localized, i.e. decay
exponentially in space \cite{loc1d}. The same statement holds for
two dimensional systems without spin-orbit interactions \cite{loc2d}.  
By contrast the single-particle states
in three dimensions are either localized or extended as a result of
the Anderson transition \cite{Anderson}. In this paper we demonstrate that the 1D
gas of weakly short-range interacting bosons in the presence of disorder exhibits
a finite temperature phase transition between two distinct states,
{\it fluid} and {\it insulator}, and the transition temperature
depends on the disorder. None of these states is characterized by a
long-range order or long-range spatial correlations. Moreover,
thermodynamic functions, such as specific heat, do not have
singularities at the transition point. From this point of view, the
dogma is not violated. Nevertheless, this is a true albeit
non-conventional phase transition, because transport and energy
dissipation properties of the fluid and insulator phases are
dramatically different and are singular at the transition.  The
difference between the fluid and insulator phases can be qualitatively
understood by comparing two many-body 1D systems: interacting
particles without disorder and the 1D Anderson insulator (disorder
without interactions). In the fluid without disorder, the dissipation
of energy of an arbitrarily slow external field and mass transport are
possible. At the same time, even at finite temperatures there is no
mass transport in Anderson insulators and the energy dissipation
vanishes for the frequency of the external field tending to zero. Here
we show that interacting 1D bosons in disorder demonstrate one of the
two types of behavior and describe the phase diagram in the
temperature-disorder plane.  We thus provide an answer to the subtle
question of how the interaction between disordered particles
may suppress Anderson localization and permit them to acquire the fluid behavior. This was the key problem for  
charge transport in electronic systems, and it is now emerging in
a new light in the studies of disordered ultracold bosons. These
studies are driven by fundamental interest and by potential
applications of atom waveguides on a chip \cite{waveguides}. Recent remarkable
experiments \cite{Aspect,Inguscio} have demonstrated Anderson localization in expanding
extremely dilute quasi-1D Bose gases, and the investigations of effects of
the interparticle interaction in relatively dense clouds are underway.

\section{Single particle localization in 1D}

\begin{widetext}

\begin{figure}[h]

\vspace*{-1cm}

\includegraphics[width=0.9\textwidth]{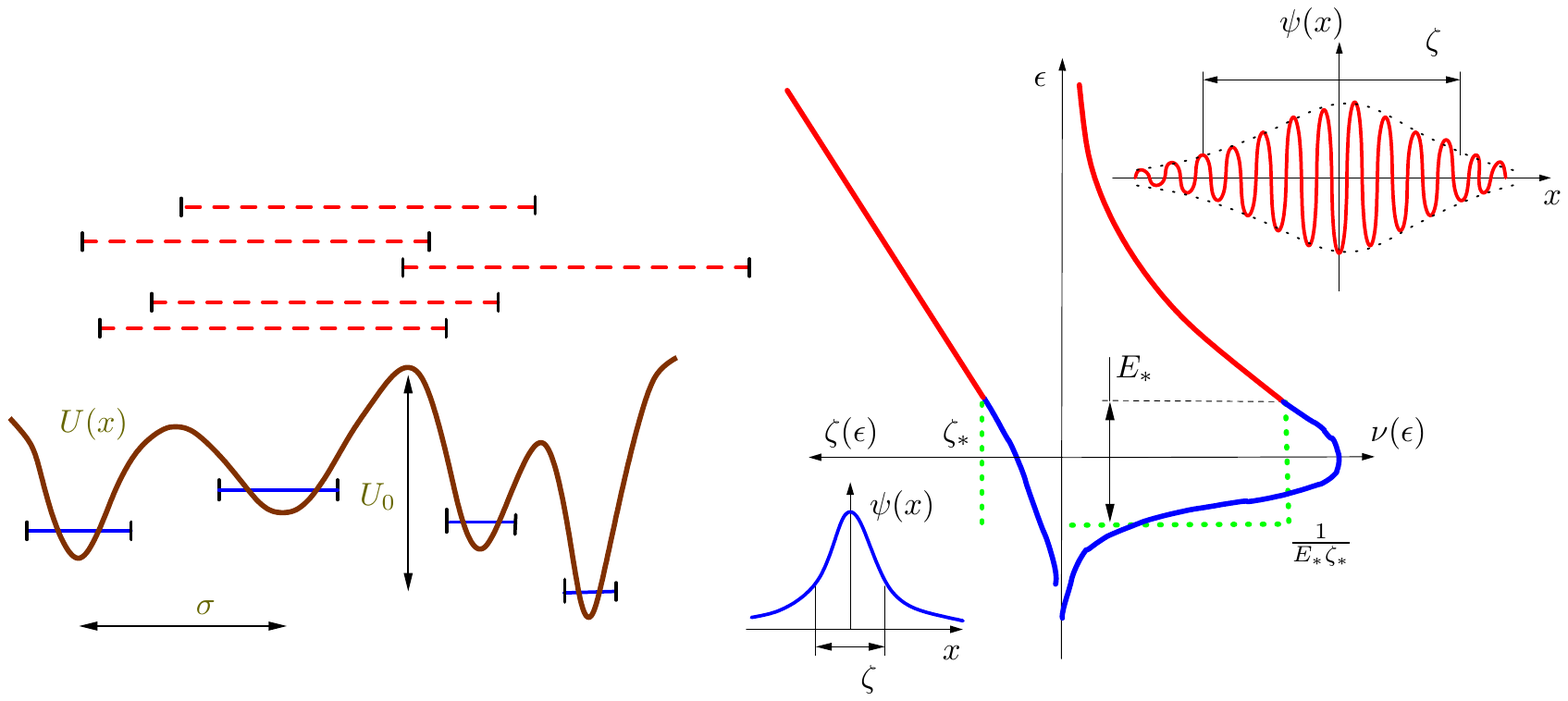}

\vspace*{-15cm}

\caption{Properties of single particle localization. 
In (a) the brown curve shows the disorder potential $U(x)$, with blue solid lines and dashed red lines indicating the location
  of tail and high-energy states, respectively. The shape of the
  wavefunctions of these states is shown by blue (tail) and red
  (high-energy) curves $\psi(x)$. In (b) the density of states $\nu$
  and localization length $\zeta$ versus energy $\epsilon$ are shown
  in blue for the DoS peak and tail states, and in red for high-energy
  states. For studying the many-body localization
  transition, $\nu(\epsilon)$ and $\zeta(\epsilon)$ in the DoS peak (low-energy states) can be approximated by
  the green dotted lines. \label{fig1}}
\end{figure}
\end{widetext}

Let us first discuss the density of states (DoS) for a single particle with
mass $m$ in 1D and introduce relevant energy and distance scales.  In
the absence of disorder the particle eigenstates are plane waves with
energies $\epsilon>0$. The DoS is
$\nu_0(\epsilon)=\sqrt{m/2\pi^2\hbar^2\epsilon}$ and it diverges at
$\epsilon\rightarrow 0$. What happens in a static random potential? At large and positive
$\epsilon$ the DoS is only slightly affected by disorder:
$\nu(\epsilon)\approx\nu_0(\epsilon)\sim 1/\sqrt{\epsilon}$. At the
same time, the disorder cuts off the DoS divergence at $\epsilon=0$
and transforms it into a peak with a finite height $\nu_*$ and width
$E_*$ as shown in Fig.~\ref{fig1}. Also, the disorder creates states with
negative energies, thus making the DoS finite at $\epsilon<0$. For
large negative $\epsilon$ the DoS is exponentially small, and this part
of the curve $\nu(\epsilon)$ is known as Lifshitz tail \cite{Lifshitz, Halperin,Zittarz}.

Both $\nu_*$ and $E_*$ are determined by the statistics of the random
potential $U(x)$. For simplicity we assume a short-range
Gaussian potential with the amplitude $U_0$ and correlation
length $\sigma$ such that condition $U_0\ll\hbar^2/m\sigma^2$ holds\cite{footnote}. 
Then the only 
relevant energy and length scales are \cite{Lifshitz,Halperin,Zittarz}:
\begin{eqnarray}
&&E_*\sim(U_0^4\sigma^2m/\hbar^2)^{1/3};   \label{Estar}\\
&&\zeta_*=\hbar/\sqrt{mE_*}=(\hbar^4/U_0^2\sigma m^2)^{1/3}.   \label{zetastar}
\end{eqnarray}
They determine the width of the DoS peak and the maximum DoS value
$\nu_*\sim1/E_*\zeta_*$. In order to obtain these scales consider a
weakly bound state of a particle in the potential $U(x)$, with an
extension of the wave function, $\zeta\gg\sigma$. The particle energy
can be written as $E\sim
(\hbar^2/2m\zeta^2\,-U_0\sqrt{\sigma/\zeta})$, where for the Gaussian
disorder the potential energy term is obtained multiplying the
contribution of each potential well, $U_0\sigma/\zeta$, by
$\sqrt{\zeta/\sigma}$, which is the square root of the number of wells
on the length scale $\zeta$. The energy $E$ reaches a minimum value
$E_*$ at $\zeta\sim\zeta_*$, with $E_*$ and $\zeta_*$ given by
Eqs.~ (\ref{Estar}) and (\ref{zetastar}).

We thus see that the single-particle spectrum can be divided into
three parts: high-energy states with $\epsilon\gg E_*$ and
$\nu(\epsilon)\sim\sqrt{m/\hbar^2\epsilon}$, low-energy states
located in the region of the DoS peak, and Lifshitz tail at negative
energies. As we already noted, all single-particle eigenfunctions in 1D
are localized with an energy-dependent localization length
$\zeta(\epsilon)$. For high-energy states, $\epsilon \gg E_*$, we have
$\zeta(\epsilon)\sim\epsilon\zeta_*/E_*$, whereas in the Lifshitz tail
$\zeta(\epsilon)\sim\zeta_*\sqrt{E_*/|\epsilon|}$ \cite{Lifshitz,Halperin,Zittarz}.  For the
low-energy states in the DoS peak,  $|\epsilon| \lesssim E_*$,  the localization length is
$\zeta(\epsilon)\sim\zeta_*$.

\section{Many-body localization-delocalization transition}

Repulsive short-range interaction between bosons  in 1D gives rise to two other energy scales in
addition to $E_*$: the temperature of quantum degeneracy
$T_d=\hbar^2n^2/m$ (we use the units with Boltzmann constant $k_B=1$),
and the interaction energy per particle, $ng$,
where $g$ is the coupling constant for the interaction,
and $n$ is the mean boson density. We focus on the weakly
interacting regime where the dimensionless coupling strength is small:
\begin{equation}    \label{gamma}
\gamma\equiv ng/T_d\ll 1
\end{equation}
so that at the mean separation between the bosons their wavefunction is
not influenced by the interactions.  
It is convenient to introduce the dimensionless temperature
\begin{equation}
t=\frac{T}{ng}=\frac{1}{\gamma}\frac{T}{T_d}.
\label{t}
\end{equation}
Another dimensionless parameter
characterizes the strength of the disorder:
\begin{equation}      \label{kappa}
\kappa\equiv E_*/ng
\end{equation}
so that large values of $\kappa$ correspond to strong disorder.  

At this point we should make two important statements. If the disorder is extremely
strong ($\kappa\rightarrow\infty$), the bosons occupy only
states in the Lifshitz tail. As the DoS in the tail
is exponentially small, the bosons are distributed
among the "lakes" located exponentially far from each
other. The bosons can not hop between the lakes, and
the system is in an insulating state.  The reduction of $\kappa$
at $T=0$ eventually (for $\kappa=\kappa_c\simeq 1$) transforms the insulator into an algebraic
superfluid (spatial phase correlations do decay, but only
algebraically). This 
Kosterlitz-Thouless type transition was first analyzed in Ref.~\cite{Giammarchi} and more recently
discussed in relation to disordered Josephson chains \cite{Rafael} and to
cold atomic gases \cite{nat,maciej}. However, well before the
transition from insulating to superfluid state
most of the particles find themselves in low-energy states where the
DoS is much higher than in the Lifshitz tail. Thus, we may neglect 
the tail in our discussion of the fluid-insulator
transition and consider only low-energy and high-energy states (green
and red in Fig.~\ref{fig1}b).

Second, although the interaction between bosons renormalizes (screens)
the disordered potential $U(x)$, this does not change the
picture of single-particle eigenstates in our discussion. The reason
is that relevant particle energies are of the order of $E_*$ or
larger. For $\kappa\gtrsim 1$ they exceed the interaction energy $ng$,
and as we will see there is no need to consider 
$\kappa\ll 1$. In this respect, the main effect of the interaction on
the ground state of the system is not screening the random potential
but rather controlling the occupation of single-particle
states.

\begin{widetext}

\begin{figure}[h]

\vspace*{-1cm}

\includegraphics[width=0.9\textwidth]{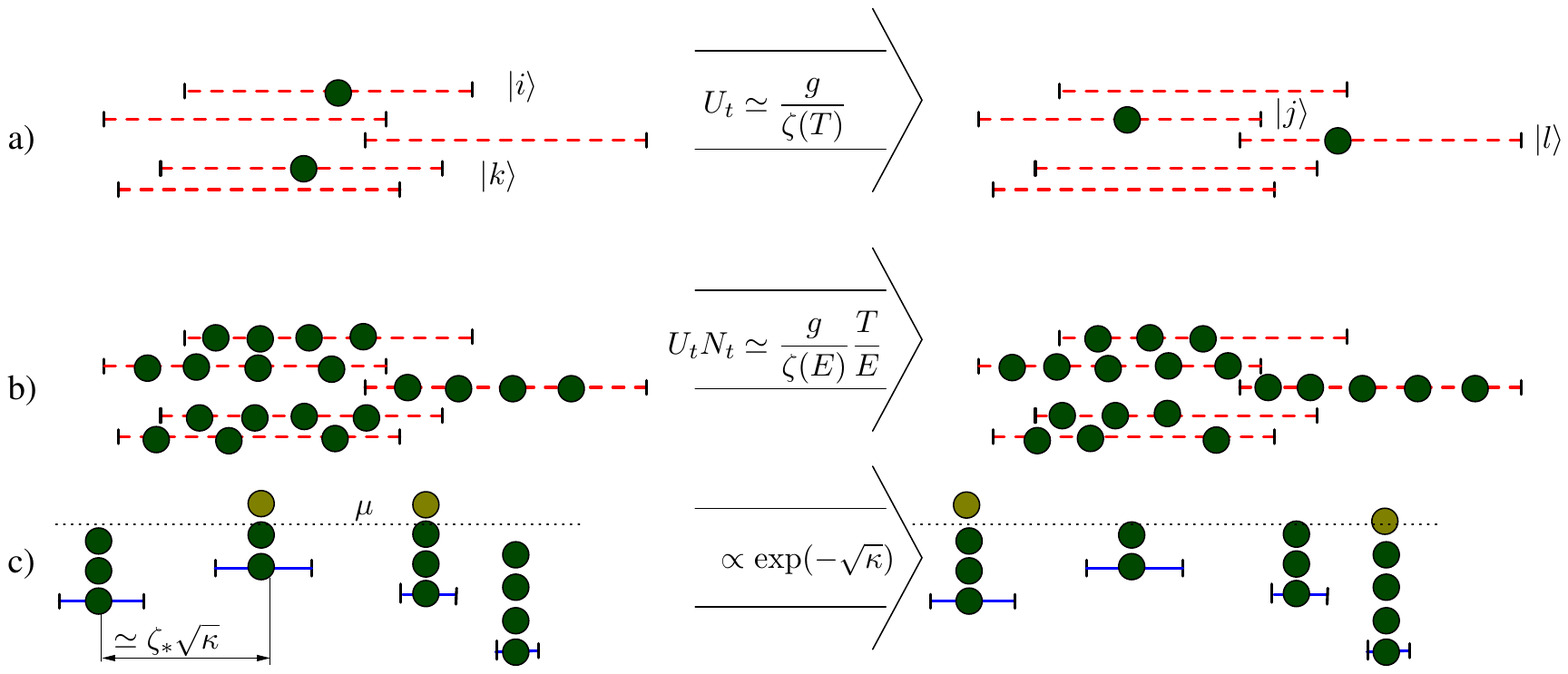}

\vspace*{-15cm}

\caption{Scattering processes leading to the many-body localization-delocalization transition for a classical Bose gas ($T>T_d$) in (a), for a degenerate thermal Bose
gas ($T_d\sqrt{\gamma}<T<T_d$) in (b), and for the low-temperature regime ($T<T_d\sqrt{\gamma}$) in (c).
\protect\label{fig2}}
\end{figure}
\end{widetext}

At a finite temperature $T$ it is crucial to take into account
two-body processes that change occupation numbers and can dramatically
affect the properties of the system. For example, the system of
interacting localized fermions (electrons) can have a finite DC
conductivity even in the absence of coupling with any outside bath
\cite{BAA}, whereas without interactions the conductivity is exactly zero
at any $T$. In the presence of the interactions the conductivity
remains zero unless the temperature exceeds a critical value
$T_c$. This transition can be thought of \cite{BAA} as Anderson
localization \cite{Anderson} of many-body wavefunctions. The critical temperature
$T_c$ depends on the interaction strength. This is the many-body
analog of the mobility edge \cite{mobilitythreshold1,mobilitythreshold2} that separates bands of localized and
extended states in the single-particle Anderson transition.

The Anderson transition is based on the fact that two quantum states
belonging to different lattice sites hybridize provided that the
hopping matrix element between these states exceeds the difference in
onsite energies. As soon as the density of the hybridized pairs
exceeds a critical value the eigenstates turn out to be extended
\cite{Anderson}. The many-body localization-delocalization transition can be
qualitatively understood by extending this physical picture to states
of more than one particle. 

Consider an occupied localized
single-particle state $|i\rangle$ with energy $\epsilon_i$. The
interaction of a particle occupying this state with a particle in the
state $|k\rangle$ can transfer the $|i\rangle$-state particle to the
state $|j\rangle$, transferring simultaneously the $|k\rangle$-state
particle to another state $|l\rangle$ (see Fig.~\ref{fig2}). Let $U_{ik,jl}$ be
the matrix element of this process. Due to the exponential decay of
the localized wave functions one may assume that $U_{ik,jl}=0$ unless
all four states are located near each other. Moreover, it turns out \cite{matrixelements1,matrixelements2} 
 that the matrix element rapidly decreases with an increase in the
energy transfer $|\epsilon_i-\epsilon_k|$. We thus may confine
ourselves to the case where the states $|i\rangle$ and $|k\rangle$ are
nearest neighbors in the energy space. For simplicity we replace
$U_{ik,jl}$ by a certain typical value $U_t$ provided that the states
$|i\rangle,|j\rangle,|k\rangle,|l\rangle$ are localized nearby and
pairwise are nearest neighbors in energy.

In a random system the energies of the final and initial states can
not be matched exactly. As long as
the energy mismatch
$\Delta_{ik,jl}=|\epsilon_i+\epsilon_k-\epsilon_j-\epsilon_l|$ exceeds
$U_{ik,jl}$ the effect of the interactions on the quantum state of 4
particles is negligible. One may say that single-particle excitations
do not decay \cite{AGKL}. Suppose that the interaction is weak and a
typical mismatch $\Delta_t$ exceeds the matrix element $U_t$. Does it
mean that single-particle excitations have an infinite lifetime? The
answer depends on the number of channels, $N_1$, for the decay of a given excitation (more precisely, $N_1$ is
the number of possible processes
$|i\rangle,|k\rangle\rightarrow|j\rangle,|l\rangle$ that involve a
given state $|i\rangle$). Indeed, with probability of order unity,
these processes should have a channel with the mismatch that is
smaller by a factor of $N_1$ than the typical value
$\Delta_t$. Therefore, $U_t$ should be compared with
$\Delta_t/N_1$. Note that $N_1$ plays a role of the number of nearest
neighbors in the single-particle localization problem. However, in the
many-body case $N_1$ is determined by the density of thermal
excitations and is temperature dependent. Since
characteristic single-particle energies and localization lengths
are determined by the temperature, both $\Delta_t$ and $U_t$ are also 
temperature dependent. As a result, there is a
critical temperature $T_c$ following from the equation
\begin{equation}      \label{critcond}
\Delta_t(T_c)=U_t(T_c)N_1(T_c).
\end{equation}
At $T>T_c$ many-body states are extended, i.e. they are linear
combinations of one-, two-particle, etc. excitations, and the
number of terms is infinite. This leads to the fluid
behavior. For $T<T_c$ the many-body localization takes
place and one expects the insulating behavior.

Note that the arguments given above and Eq.~(\ref{critcond}) are
general and independent of quantum statistics of the particles. At the
same time, both $U_t(T)$ and $N_1(T)$ do depend on the statistics. 
For disordered bosons these quantities are determined by the
density of single particle states $\nu(\epsilon)$ and by the
occupation number $N(\epsilon)$. The latter is controlled by the
chemical potential $\mu$ which is related to the mean density $n$ and
temperature $T$ by the normalization condition $n=\int d\epsilon
N(\epsilon)\nu(\epsilon)$, with $N(\epsilon)$ determined by the
Bose-Einstein distribution. More precisely, as long as the interaction
is weak ($\gamma\ll 1$) the occupation is
$N(\epsilon)=\{\exp[(\epsilon_{HF}-\mu)/T]-1\}^{-1}$, where
$\epsilon_{HF}$ differs from the single-particle energy $\epsilon$ by
the Hartree-Fock corrections. We will see that these corrections
become important only at sufficiently low $T$ and are negligible at
high temperatures where in the vicinity of the transition relevant particle energies
 greatly exceed $ng$.

\section{Phase diagram}

We now use Eq.~(\ref{critcond}) and analyze the phase diagram of
weakly interacting disordered 1D bosons. It is convenient to represent
the phase diagram in terms of the dimensionless temperature $t=T/ng$
and find the dependence of $t_c$ on the strength of
the disorder $\kappa$. The relation $t=t_c(\kappa)$ determines the
boundary between the fluid and insulator phases in the $(t,\kappa)$
plane. Alternatively, one can  speak of a temperature-dependent
critical disorder $\kappa_c(t)$.

At high temperatures $T>T_d$, or $t>\gamma^{-1}$, the Bose gas is not
degenerate and a characteristic energy of particles is of the order of
$T\gg ng$. As will be seen, near the transition temperature we have
$E_*<T$ so that most of the particles are in the high-energy
states and the occupation of all the states is small and described bu
Boltzmann distribution. In this case
the typical matrix element $U_t(T)$ in Eq.~(\ref{critcond}) does not
depend on the occupation and can be estimated as $U_t(T)\sim
g/\zeta(T)$. The typical mismatch is the meen nearest neighbor
energy spacing: $\Delta_t\sim[\nu(T)\zeta(T)]^{-1}$. The quantity
$N_1(T)$ is given by the number of particles localized
within a distance $\sim\zeta(T)$ from a given state $|i\rangle$,
i.e. $N_1(T)\approx\zeta(T)n$. Then, Eq.~(\ref{critcond}) is
reduced to $ng\nu(T_c)\zeta(T_c)\sim 1$. Using the high-energy
localization length $\zeta(T)=\zeta_* T/E_*$ and the density of states
$\nu(T)=1/\sqrt{E_*\zeta_*^2 T}$, with the help of Eqs.~(\ref{t})
and (\ref{kappa}) we obtain the high temperature relation
for the critical disorder :
\begin{equation}           \label{highTcond}
\kappa_c(t)\sim t^{1/3};\,\,\,\,\,\,\,\,t>\gamma^{-1}.
\end{equation}
Close to the transition we have $E_*/T=\kappa/t\sim
t^{-2/3}\ll\gamma^{2/3}\ll 1$, which justifies our initial assumption
that most of the particles are in the high-energy states.

As the temperature is reduced below $T_d$ the Bose gas becomes
degenerate. In the absence of disorder the chemical
potential is $\mu=-T^2/T_d$ as long as $T>T_d\sqrt{\gamma}$, or
$t>1/\sqrt{\gamma}$. Characteristic energies of particles are $\sim
|\mu|<T$. However, as will be seen below, they still exceed both $ng$ and $E_*$. Therefore,
most of the 1D disordered interacting bosons in the temperature
interval $\gamma^{-1/2}<t<\gamma^{-1}$ occupy high-energy states and
one can use the ideal Bose gas distribution $N(\epsilon)$,
i.e. neglect Hartree-Fock corrections to single-particle energies. The
major part of particles has energies $\epsilon\lesssim|\mu|$ and is
characterized by a multiple occupation $N(\epsilon)\sim T/\epsilon>
1$. This manifests itself in the dependence of interaction matrix
elements on $N(\epsilon)$.

In this regime there are two energy scales ($T$ and $\mu \ll T$)
characterizing the distribution of particles. What are the particle energies
that determine the many-body delocalization? Let us apply Eq.~(\ref{critcond}) to
particles with energies $\epsilon\sim E$ in the energy interval of
width $\sim E$. A typical value of the energy spacing is
$\Delta_t=1/\nu(E)\zeta(E)$, and the typical matrix element of the
two-body interaction is enhanced due to a multiple occupation of
single-particle states: $U_t=[g/\zeta(E)](T/E)$. The
number $N_1$ of occupied levels at distances smaller than $\zeta(E)$
from a given state is $N_1(E)\sim E\nu(E)\zeta(E)$ and Eq.~(\ref{critcond}) takes the form
$$N_1U_t/\Delta_t\sim gT_c\nu^2(E)\zeta(E).$$
Then, using the high-energy density of states
$\nu(E)=1/\sqrt{E_*\zeta^2E}$ and localization length
$\zeta(E)=\zeta_*E/E_*$ we find a remarkable result: the criterion of
delocalization does not involve the single-particle energy scale! The
transition temperature follows from the relation $gT_c\sim
E_*^2\zeta_*=\sqrt{\hbar^2E_*^3/m}$. (The fact that this is valid for
all energy scales suggests that the expression for the critical
temperature/disorder can contain a prefactor logarithmic in $\gamma$,
which we neglect). In terms of the parameters $t$ and $\kappa$ the
relation for the critical disorder becomes 
\begin{equation}         \label{intTcond}
\kappa_c=t^{2/3}\gamma^{1/3};\,\,\,\,\,\,\,\,\gamma^{-1/2}<t<\gamma^{-1}.
\end{equation}
Equation (\ref{intTcond}) shows that $\kappa_c\gtrsim 1$ in the entire temperature interval $\gamma^{-1/2}<t<\gamma^{-1}$. Hence, close to the transition we have $E_*\gtrsim ng$
and characteristic particle energies are $\sim |\mu|\gtrsim E_*\gtrsim ng$. This justifies our assumption that the major part of particles occupies high-energy states, and the
interparticle interaction affects neither the occupation numbers $N(\epsilon)$ nor the chemical potential $\mu$.
   
Consider now $T=0$. For $\kappa\gg 1$ the boson density is fragmented
into "lakes". Lake number $i$ is formed by $N_i$ bosons in the
single-particle eigenstate $ |i\rangle$ which is characterized by
energy $\epsilon_i$ and localization length $\zeta_i\approx\zeta_*$.
The energy cost $E_i$ of bringing an extra particle to this lake is
enhanced by the repulsive interaction between the bosons, $E_i\approx
\epsilon_i+gN_i/\zeta_*$, and it should be equal to the global
chemical potential $\mu$ measured from the lowest low-energy state
$\epsilon=0$. The bosons thus occupy only states below the chemical
potential, $\epsilon_i<\mu$, with the occupation numbers
$N_i\approx(\mu-\epsilon_i)\zeta_*/g$, and as long as $\mu<E_*$ only
low-energy states are occupied. The density of these states is
$\nu_*=1/E_*\zeta_*$, and thus the mean density of bosons is related
to the chemical potential by $n=\mu^2/2gE_*$. The chemical potential
can be expressed in terms of the parameter $\kappa$ as
$\mu=E_*/\sqrt{\kappa}$, i.e. the chemical potential is indeed smaller
than $E_*$ provided that $\kappa>1$. In this regime only a small
fraction ($\sim\mu/E_*$) of low-energy states is occupied and
neighboring lakes are separated by a distance
$l(\kappa)\sim\zeta_*\sqrt{\kappa}$, while their size is
$\sim\zeta_*$. Using equations (\ref{Estar})-(\ref{kappa}) one can
show that a typical lake ($\epsilon_i\sim\mu$) contains $N_i\sim
n\zeta_*\sqrt{\kappa}=1/\sqrt{\gamma}$ bosons.

The fact that $l(\kappa)\gg\zeta_*$ implies that for $\kappa\gg 1$ the
system is a strong insulator: the coupling between different lakes is
exponentially small in $\kappa$. As soon as $\kappa$ is reduced to the
value of the order of unity, the distance between neighboring lakes
becomes of the order of their size $\zeta_*$ and the interlake
coupling drives the system to the fluid state. So, the
insulator-fluid Berezinskii-Kosterlitz-Thouless
transition at $T=0$ occurs for $\kappa_c\sim 1$.
        
Note that at the lower bound of the temperature interval in
Eq.~(\ref{intTcond}), $t\sim\gamma^{-1/2}$, the critical disorder is also
$\kappa_c\sim 1$. Therefore, one expects that in the
entire temperature range $t<\gamma^{-1/2}$ it remains $\kappa_c\sim 1$.

Why the insulating state of bosons is stable at these temperatures as
long as $\kappa>1$, i.e. $E_*>ng$? Let the disorder be as strong and
reduce the temperature below $\sqrt{E_*T_d}$
($t<\sqrt{\kappa/\gamma}$). Under these conditions we have $|\mu|<E_*$
and only a few bosons are hosted by high-energy states. For high-energy 
bosons ($\epsilon>|\mu|$) the condition (\ref{critcond}) is not
satisfied because their density is too small for the many-body
delocalization. It turns out that the main body of the bosons, which
occupy low-energy states, also forms an insulator. Indeed, for
low-energy states we have $\epsilon<E_*$ and the number of channels
$N_1\sim\epsilon\nu(\epsilon)\zeta(\epsilon)$ is smaller than unity,
since $\nu(\epsilon)\zeta(\epsilon)\sim E_*^{-1}$ . In other words,
most of the particles occupy single-particle states which are
separated from each other by distances exceeding the localization
length $\zeta_*$. This causes exponential reduction of $U_t$ and,
according to Eq.~(\ref{critcond}), prevents delocalization. 
Finite temperature in this situation does not lead to any increase in
the phase volume for available transitions. By contrast temperature 
fluctuations of the number of particles in each lake
lead to a {\em growth} of the energy mismatch by an amount $\sim
{\sqrt{Tg/\zeta_*}}$, which further suppresses the probablity to hybridize several states. 
The physical situation is somewhat similar to the one described in
Ref. \cite{OH}, where it was demonstrated numerically that in a
finite-width band with less than one state per localization length the
insulator remains stable with respect to the interactions at arbitrarily
high temperatures.

If the disorder is weak, $\kappa<1$, the chemical potential is
determined by the interaction and always exceeds $E_*$. As a result,
we have $N_1(T)>1$. The condition (\ref{critcond}) then indicates that
the insulator is unstable, and one deals with the fluid state.

The arguments presented above indicate that the phase transition line has to
be almost horisontal in the region $t <1/\sqrt{\gamma}$. We can also 
realize that at $t\to 0$ the line $\kappa_c(t)$ should terminate at
the quantum  phase transition point $\kappa=1$. Indeed, assuming
$\kappa_c(t=0)<1$ we arrive at a contraidiction as the underlying superfluid phase at
$T=0$ has delocalized excitations (phonons) at low energies. On the
other hand, the assumption of $\kappa_c >1$ is also not consistent with the
criterion (\ref{critcond}) as all of the excitations are localized
and the temperature should be finite to provide a finite density of the excitations. 


\begin{figure}[h]

\vspace*{0.2cm}

\hspace*{-1.9cm}
\includegraphics[width=2.1\columnwidth]{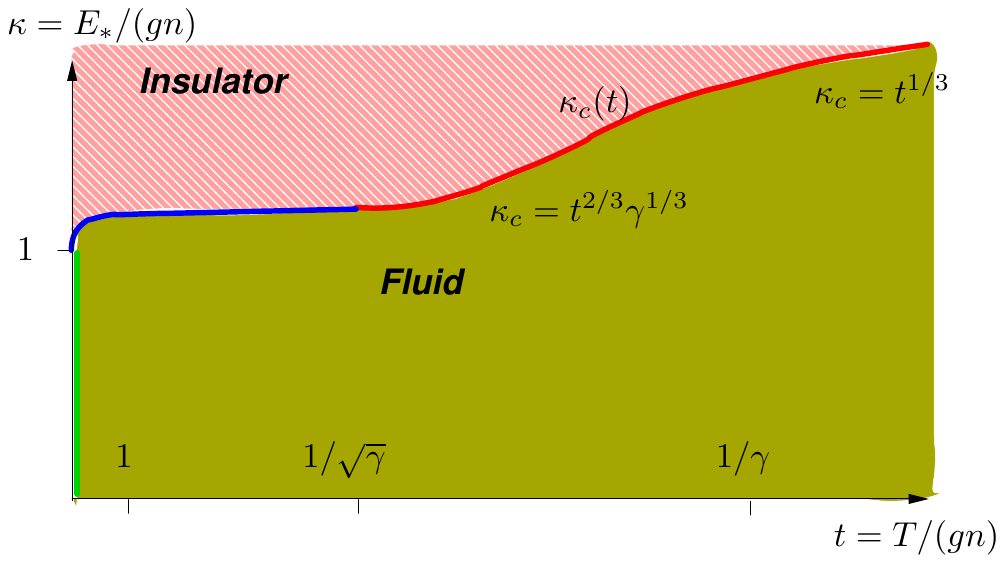}

\vspace*{-20cm}

\caption{Phase diagram for weakly interacting disordered bosons. The green line shows the zero temperature algebraic superfluid. The red part of the curve $\kappa_c(t)$ follows
from Eqs.~(\protect\ref{highTcond}) and (\protect\ref{intTcond}), and the blue part is an estimate for the low-temperature regime. 
\protect\label{fig3}}
\end{figure}

This completes our description of the finite temperature
fluid-insulator phase transition for 1D interacting bosons. The phase
diagram is presented in Fig.~\ref{fig3}. Detailed calculations and a more
accurate description of the temperature range $t<\gamma^{-1/2}$ (blue part of the curve $\kappa_c(t)$ in Fig.~\ref{fig3}) will be
presented elsewhere. 


\section{Dynamics of expansion}

The observation of the fluid-insulator phase transition described above is
feasible in experiments with cold bosonic atoms in the 1D geometry. 
For electrons in solids one measures e.g. the DC conductivity.
In quantum (neutral-atom) gases the analysis of transport properties 
is based  of the dynamics induced by significant 
external perturbations which may drive the system far from the initial state. 
We believe that the localization-delocalization transition can be identified in
the dynamics of expansion of disordered bosonic clouds released from 
the superimposed trapping potential.   

In an array of harmonically trapped quasi-1D tubes with about a hundred of atoms per tube, the density $n$ can be made $\sim 10^4$ cm$^{-1}$ so that the length $L$ of each tube is $\sim 100$ $\mu$m and the temperature of quantum degeneracy $T_d$ is of the order of tens of nanokelvins. Tuning the interaction strength by Feshbach resonances or by variations of the tight transverse confinement one can achieve the interaction energy $ng$ of the order of nanokelvins or even smaller and make $\gamma\sim 10^{-1} - 10^{-3}$. For the correlation length of the disorder $\sigma\simeq 0.3$ $\mu$m as in the experiment \cite{Aspect} and typical values of $U_0$ \cite{Aspect,Inguscio}, the localization length $\zeta_*$ is in the range of $1 - 10$ microns and the energy $E_*\gtrsim ng$. The dimensionless strength of the disorder, $\kappa$ (\ref{kappa}), ranges from about unity to large values and one can to study all temperature regimes of the phase diagram in Fig.~\ref{fig3}. Note, however, that the conditions described here are quite different from those in the experiment \cite{Aspect}, where the interaction energy was greatly exceeding $E_*$ and the system was deeply in the fluid state. 

\begin{figure}

\vspace*{-1cm}

\hspace*{-1.9cm}
\includegraphics[width=1.8\columnwidth]{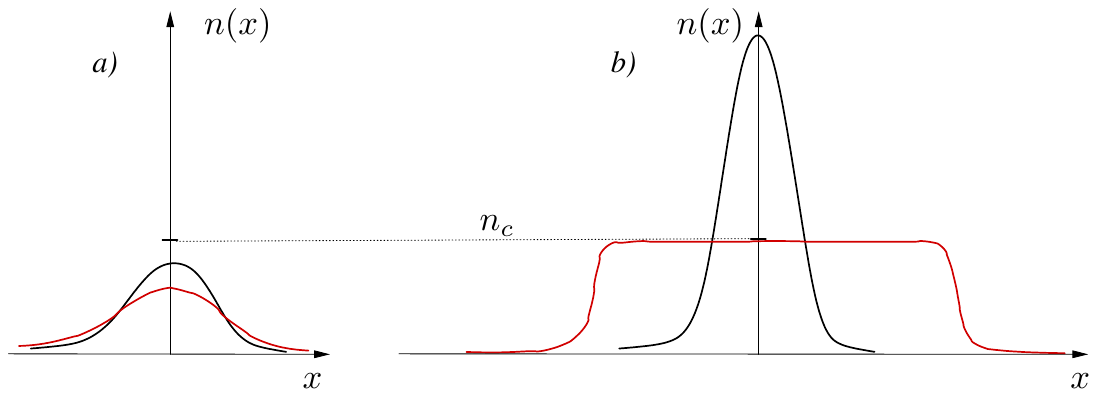}

\vspace*{-17cm}

\caption{Manifestation of the many-body localization-delocalization
  transition in the expansion of a quasi-1D cloud.  In (a) the 
  the entire initial cloud is an insulator, whereas in (b) initially
  the central part is in the fluid and the outskirts in the insulating
  phase. Initial and final shapes of the cloud are shown in black and red, 
  respectively. \protect \label{fig5}}
\end{figure}

Switching off abruptly the 1D trap but still keeping the disorder (and
the transverse tight confinement) like in the experiments \cite{Aspect,Inguscio}, is
expected to cause the expansion of the cloud. Not very far from the fluid-insulator transition, 
the localization length $\zeta$ of single-particle states in the initial cloud is 
much smaller than its size $L$. Under this condition, the size rapidly
increases by an amount of $\sim\zeta \ll L$. If the
entire initial cloud is in the insulator phase, i.e.
locally $\kappa>\kappa_c(t)$, the expansion then stops (see Fig.~\ref{fig5}a). 
If the central part of the cloud is in the fluid phase (while
the outskirts are insulators), two-body scattering processes induce further expansion of the central part.
In this stage the expansion is a slow diffusive process governed by the diffusion
equation:
\begin{equation}          \label{diffusioneq}
\frac{\partial n}{\partial \tau}=\frac{\partial}{\partial
  x}D(n,t)\frac{\partial n}{\partial x},
\end{equation}
where $\tau$ denotes the time and $t$ is the dimensionless temeperature (\ref{t}). The diffusion coefficient $D(n,t)$ strongly depends on the local density $n(x)$ and vanishes at the border of the insulating phase, i.e. for the density $n_c(t)=E_*/(g\kappa_c(t))$. Hence, the local decrease of the density stops when it becomes equal to $n_c(t)$. Thus, the density profile of the expanded cloud represents a plateau with $n=n_c(t)$ in the
central part and the initial insulating wings with $n(x)<n_c$ \cite{Shklovskii} (see Fig.~\ref{fig5}b). 

Moderately far from the transition, i.e. at $(n-n_c)/n_c\sim 1$, the diffusion coefficient can be estimated as follows. The occupation $N_i$ of the state $|i\rangle$ changes by $\sim 1$ on a time scale of the order of the inverse matrix element $\hbar/U_t\sim \hbar\zeta/gN_i$. The distance of a typical hop is $\sim\zeta$. Therefore, on a time scale $\tau_0\sim N_i\hbar/U_t\sim\hbar\zeta/g$ each boson will move by a distance $\sim\zeta$, and $D\sim\zeta^2/\tau_0\sim\zeta g/\hbar$. For high-energy states the localization length at typical energies $\epsilon$ is $\zeta(\epsilon)=\zeta_*\epsilon/E_*$, and the diffusion coefficient is given by $D\sim \epsilon\zeta_* g/\hbar E_*$. Locally, the decrease of the density ceases when it reaches the critical value $n_c(t)$. Thus, for the initial central density comparable with $n_c(t)$, the evolution of the cloud to the final shape of the plateau and wings (see Fig.~\ref{fig5}) requires a characteristic time $\tau_*\sim L^2/D$. Since $L\sim (\epsilon/m\omega^2)^{1/2}$, where $\omega$ is the initial trap frequency, we obtain
\begin{equation}           \label{taustar}
\tau_*\sim\frac{1}{\omega}\,\frac{\hbar E_*}{m\omega\zeta_* g}\sim\frac{1}{\omega}\left(\frac{ng}{\hbar\omega}\right)\frac{\kappa^{3/2}}{\gamma^{1/2}}.
\end{equation}              
However, in the temperature interval $T_d\sqrt{\gamma}<T<T_d$ most of the particles, which have energies $\epsilon\sim|\mu|=T^2/T_d$, and particles with $\epsilon\sim T$ will expand with different velocities. So, the redistribution of particles in the course of the expansion may become important and it can slightly modify the above estimate for the time $\tau_*$.

The time $\tau_*$ by far exceeds the time $\omega^{-1}$ of ballistic expansion of the clean thermal cloud. For example, near the lower bound of Eq.~(\ref{highTcond}), i.e. for $T\sim T_d$ and $\kappa\sim\gamma^{-1/3}$, we have $\tau_*\sim T_d/\hbar\omega^2$. For typical values $T_d\sim 50$ nK and $\omega\sim 10$ Hz this estimate yields $\tau_*\sim 1$ s. Moving to the lower bound of Eq.~(\ref{intTcond}) where $T\sim T_d\sqrt{\gamma}$ and $\kappa\sim 1$, the time $\tau_*$ reduces by a factor of $\sqrt{\gamma}$, i.e. it is of the order of $0.1$ s. As we see, these time scales are such that one can observe the evolution of the expanding cloud from the very
beginning until the final density distribution is formed. The expansion of the fluid part of the cloud
will also be slow in the low-temperature regime, $T<T_d\sqrt{\gamma}$ ($t<\gamma^{-1/2}$).
The analysis of the diffusion coefficient for this case is beyond
the scope of the present paper.


\label{experiments}

\section{Phase dagram in higher dimensions}

\begin{figure}
\includegraphics[width=0.4\textwidth]{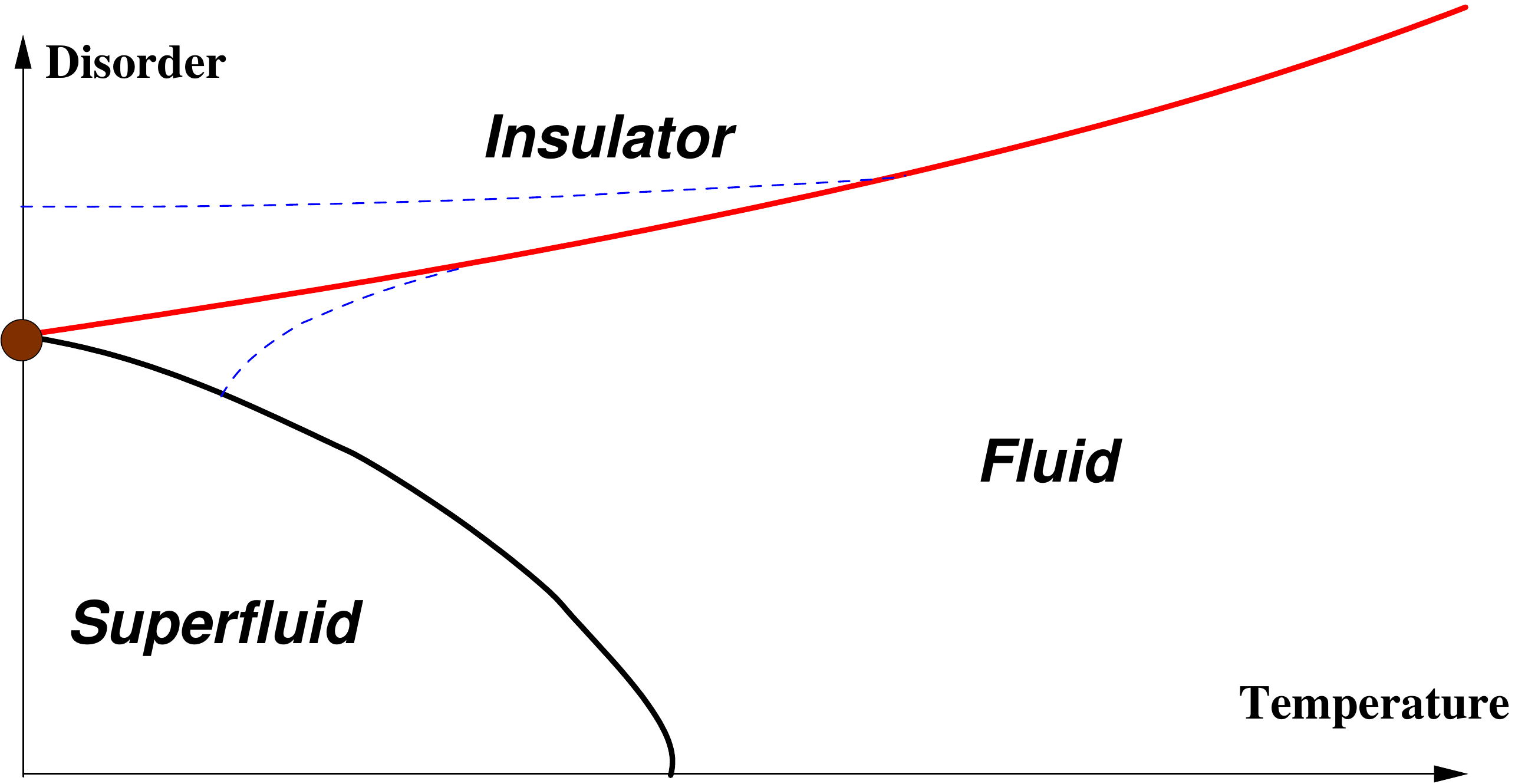}

\caption{Phase diagram for two-dimensional weakly interacting bosons.
The black curve shows the thermodynamic Berezinskii-Kosterlits-Thouless transition. The red curve
is the many-body localization-delocalization transition. The blue
dashed curves indicate phase boundaries
which can be ruled out on rather general grounds (see text).\protect\label{fig4}}
\end{figure}

We conclude our discussion of the properties of disordered interacting
bosons with a brief sketch of the phase diagram in higher dimensions. 
As is well known, in the absence of disorder bosons form a superfluid 
(algebraic superfluid in 2D) below a critical temperature ${\bar T}_c$. 
At high temperatures, $T>{\bar T}_c$, the clean system is a normal fluid. The
superfluid survives a sufficiently weak disorder, but the superfluid transition  
temperature ${\bar T}_c$ decreases with increasing the strength of the disorder 
and vanishes at a critical strength\cite{dirtybosons} (see the black curve and brown point in Fig.~\ref{fig4}).
What is the state of the disordered system at $T>{\bar T}_c$, i.e. above the black curve
in Fig.~\ref{fig4} ?  It follows from our previous
discussion that bosons can form either the normal fluid or the
insulating state. The suggested phase diagram is presented in Fig.~\ref{fig4}.

This sketch can be justified in the following way. First, it is safe
to assume that at sufficiently strong disorder bosons are in the insulating
state. The second observation is that at the critical disorder 
(brown point in Fig.~\ref{fig4}) and arbitrarily
small but finite T one should expect the normal fluid rather than the
insulator. Indeed, the zero-temperature insulator can be thought of as a system of
superfluid lakes which are separated from each other and have uncorrelated phases.
A typical size of such a lake increases with decreasing the strength of the disorder 
and diverges at the critical strength. It means that although excitations 
in the insulating phase at $T=0$ are always localized, their localization length can
be arbitrarily large. Therefore, at any finite temperature and
interaction strength there is a vicinity of the critical disorder,
where the insulator is unstable with respect to the many-body
delocalization. As a result, the insulator - normal fluid phase boundary
can not follow the lower blue curve on Fig.~\ref{fig4}.

On the other hand, the normal fluid can not be stable at $T=0$ . Indeed,
the wavefunctions of low-energy single-particle states have to be 
localized, otherwise particles can not avoid Bose-Einstein condensation 
and the system would become superfluid. At extremely low but finite temperatures, 
the density of thermal excitations is vanishingly small and the
interaction between them is unable to delocalize many-body wavefunctions.
This rules out the phase boundary following the upper blue curve in
Fig.~\ref{fig4}. Thus, the only possible option is represented
by the red solid curve \cite{comment}.

The arguments given above apply to both 2D and 3D cases. However there is a
big difference, since in 2D all single-particle states are known to be
localized, and the strong-disorder phase is a “true insulator”. This
means that the diffusion constant is zero even at finite
temperatures. At the same time, in three dimensions high-energy states can be
extended. As soon as the extended states appear they can host thermal
excitations and thus allow exponentially small but finite
diffusion. From this point of view, there is no qualitative difference
between the insulator and normal fluid, and the red curve in 3D represents a
crossover rather than a true phase transition.

\section{Concluding remarks}

A remarkable possibility to compare disordered interacting bosons with non-interacting ones is offered by a $^7$Li atomic gas, where the coupling strength $g$ can be varied by a Feshbach resonance from practically zero to large ppositive values \cite{Hulet2}. In the 1D case, achieving the strongly interacting regime where $\gamma\gtrsim 1$, is expected to present new transparent physics. For $\gamma\rightarrow\infty$ as in the recent cesium experiment \cite{Nagerl}, the bosons become impenetrable and show a clear analogy with non-interacting fermions. All single-particle states are then localized irrespective of the strength of the disorder. At intermediate values of $\gamma$ one expects a peculiar interplay between the interparticle interaction and temperature. For $T\gg ng$ the situation should be the same as desribed in Section III for the high-temperature regime, $T\gg T_d$, and in this sense equation (\ref{highTcond}) is universal. In the other extreme, $T\lesssim ng$, an increase in the interaction strength first leads to the localization-delocalization transition at small $\gamma$, but then causes a reentrance to the insulating phase at a critical disorder-dependent value of $\gamma$. This behavior is expected because at $T=0$ even an infinitesimally small disorder leads to the appearance of an insulating (Bose glass) phase     
if $\gamma$ is sufficiently large \cite{Giammarchi}. The comparison of the insulating phases emerging at small and large $\gamma$ with each other is supposed to shed new light on the structure of Bose glasses.


\vspace*{4mm}

\section*{Acknowledgements} 
We are grateful to Alain Aspect and Jean Dalibard for
interesting discussions, and to Matthew Foster and Leonid Glazman for
comments on the manuscript. We acknowledge support from US
DOE contract No. DE-AC02-06CH11357, from the IFRAF Institute of Ile de France,
and from ANR (Grant ANR-08-BLAN-0165). G.S. was also supported by the Dutch Foundation FOM. 
Part of the work was performed during the workshop "From Femtoscience to Nanoscience: Nuclei, Quantum
Dots, and Nanostructures" in the Institute of Nuclear Theory at the
University of Washington.



\begin{thebibliography}{99}

\bibitem{textbook1} van Hove, L. Sur l'integrale de configuration pour les systemes de particules a une dimension. {\it Physica} {\bf 16}, 137-143 (1950).
\bibitem{textbook2} Landau, L.D. $\&$ Lifshits, E.M. {\it Statistical Physics} (Pergamon Press, London, 1958).
\bibitem{loc1d} Gertsenshtein, M.E. $\&$ Vasil'ev, V.B. Waveguides with random inhomogeneities and brownian motion In the Lobachevsky plane. {\it Theor. Probab. Appl.} {\bf 4}, 
391-398 (1959).
\bibitem{loc2d} Abrahams, E.,  Anderson, P.W.,  Licciardello, D.C. $\&$ Ramakrshnan, T.V. Scaling theory of Localization - Absence of quantum diffusion in 2 dimensions. {\it Phys. Rev. Lett.} {\bf 42}, 673-676 (1979).
\bibitem{Anderson} Anderson, P.W. Absence of diffusion in certain random lattices. {\it Phys. Rev.} {\bf 109}, 1492-1505 (1958).
\bibitem{waveguides} Fortagh, J. $\&$ Zimmermann C. Magnetic microtraps for ultracold atoms. {\it Rev. Mod. Phys. } {\bf 79,} 235-239 (2007).
\bibitem{Aspect} Billy, J. et al. Direct observation of Anderson localization of matter waves in a controlled disorder. {\it Nature} {\bf 453,} 891-894 (2008).
\bibitem{Inguscio} Roati, G. et al. Anderson localization of a non-interacting Bose-Einstein condensate. {\it Nature} {\bf 453,} 895-898 (2008). 
\bibitem{Lifshitz}Lifshitz, I.M. Energy spectrum structure and quantum states of disordered condensed systems. {\it Sov. Phys. Usp.} {\bf 7}, 549-573 (1965).
\bibitem{Halperin} Halperin, B.I. $\&$ Lax, M. Impurity-band tails in high density limit I. Minimum counting method. {\it Phys. Rev.} {\bf 148}, 722-740 (1966). 
\bibitem{Zittarz} Zittartz, J. $\&$ Langer, J.S. Theory of bound states in a random potential. {\it Phys. Rev.} {\bf 148}, 741-747 (1966).
\bibitem{footnote} In the opposite limit, the random potential gets screened by the interparticle interaction. The phase diagram in Fig.~\ref{fig3} is then modified quantitatively rather than qualitatively: one may replace $U(x)$ by an effective screened potential $U_{eff}(x)$. For weaky interacting bosons the fluid-insulator transition is expected in the region where the amplitude and correlation length of $U_{eff}(x)$ satisfy the inequality $U_{eff}\lesssim\hbar^2/m\sigma_{eff}^2$. This issue will be analyzed elsewhere.
\bibitem{Giammarchi} Giamarchi, T. $\&$ Schulz, H. Anderson localization and interactions in one-dimensional metals. {\it Phys. Rev. B} {\bf 37}, 325-340 (1988). 
\bibitem{Rafael}Altman, E., Kafri, Y., Polkovnikov, A. $\&$ Refael, G. Insulating phases and superfluid-insulator transition of disordered boson chains. 
{\it Phys. Rev. Lett}. {\bf 100}, 170402 (2008).
\bibitem{nat} Falco, G.M., Nattermann, T. $\&$ Pokrovsky, V.L. Weakly
interacting Bose gas in a random environment. {\it Phys. Rev. B}, {\bf 80}, 104515 (2009).
\bibitem{maciej} Lugan, P., et al. Ultracold Bose gases in 1D disorder: From Lifshitz glass to Bose-Einstein condensate. {\it Phys. Rev. Lett.} {\bf 98,} 170403 (2007). 
\bibitem{BAA} Basko, D.M., Aleiner, I.L. $\&$ Altshuler, B.L. Metal-insulator transition in a weakly interacting many-electron system with localized single-particle states. {Annals of
  Physics} {\bf 321}, 1126-1205 (2006).
\bibitem{mobilitythreshold1} Mott, N.F. $\&$ Twose, W.D. The theory of impurity conduction. {\it Advances in Physics} {\bf 10}, 107-163 (1961). 
\bibitem{mobilitythreshold2}  Mott, N.F. Conduction in non-crystalline systems: 4. Anderson localization in a disordered lattice. {\it Phil. Mag.} {\bf 22}, 7-29 (1970).
\bibitem{matrixelements1} Mirlin, A.D. Statistics of energy levels and eigenfunctions in disordered systems. {\it Phys. Rep.} {\bf 326},  259-382 (2000). 
\bibitem{matrixelements2} Aleiner, I.L., Brouwer, P.W. $\&$ Glazman, L.I. Quantum effects in Coulomb blockade. {\it Phys. Rep.} {\bf 358}, 309-440 (2002).
\bibitem{AGKL} Altshuler, B.L., Gefen, Y., Kamenev, A. $\&$ Levitov, L.S. Quasiparticle lifetime in a finite system: A nonperturbative approach.
{\it Phys. Rev. Lett.} {\bf 78}, 2803-2806 (1997).te
\bibitem{OH} Oganesyan, V. $\&$ Huse, D.A. Localization of interacting fermions at high temperature. {\it Phys. Rev. B} {\bf 75}, 155111 (2007).
\bibitem{Shklovskii} A similar shape of the expanded cloud was 
suggested for 3D disordered bosons at $T\rightarrow 0$: Shklovskii,
B.I. ``Superfluid-insulator transition in ''Dirty'' ultracold Fermi
gas''. {\it Semiconductors} {\bf 42}, 909-913 (2008). However, near
the fluid-insulator transition for a disordered
 interacting system in the limit of $T\rightarrow 0$, 
the entropy production (and heating) during the expansion becomes important, and it stops due to the many-body
localization-delocalization transition described in our paper.  
\bibitem{dirtybosons} Fisher, M.P., Weichman, P.B., Grinstein, G., $\&$ Fisher D.S. Boson localization and superfluid-insulator transition. {\it Phys.Rev. B} {\bf 40}, 546-570 (1989). 
\bibitem{comment} Our conclusion for the 2D case contradicts the recent work: 
Mueller, M. Purely electronic transport and localization in the Bose
glass, arXiv:0909.2260. This paper aussumes the existence of
delocalized states for 2D non-interacting particles in the absence of spin-orbit scattering. We do not see any basis for this assumption.
\bibitem{Hulet2} Pollack, S.E. et al. Extreme tunability of interactions in a $^7$Li Bose-Einstein condensate. Phys. Rev. Lett. {\bf 102}, 090402 (2009).
\bibitem{Nagerl} Haler, E. et al. Realizatin of an excited strongly correlated quantum gas phase. Science {\bf 325}, 1224-1227 (2009). 
\end{thebibliography}
\end{document}